\documentclass{article}
\usepackage{spconf,amsmath,graphicx,hyperref}
\usepackage{amsmath,amsfonts}
\usepackage{algorithmic}
\usepackage{algorithm}
\usepackage{array}
\usepackage[caption=false,font=normalsize,labelfont=sf,textfont=sf]{subfig}
\usepackage{textcomp}
\usepackage{stfloats}
\usepackage{url}
\usepackage{verbatim}
\usepackage{graphicx}
\usepackage{cite}
\usepackage{hyperref}
\usepackage{times}
\usepackage{latexsym}
\usepackage{amssymb}
\usepackage{pifont}
\usepackage{booktabs} 
\usepackage{threeparttable} 

\usepackage{setspace}
\usepackage{multirow}
\usepackage[table,xcdraw]{xcolor}
\usepackage[T1]{fontenc}

\usepackage{caption}    
\usepackage{siunitx}    

\usepackage[utf8]{inputenc}

\usepackage{microtype}

\usepackage{makecell}   

\usepackage{inconsolata}

\usepackage{graphicx}

\usepackage{xcolor}

\definecolor{nightblue}{RGB}{25, 25, 112} 
\definecolor{orangered}{RGB}{255, 69, 0}  

\usepackage{amsmath,amsfonts}
\usepackage{algorithmic}
\usepackage{algorithm}

\title{InstructAudio: Unified speech and music generation with natural language instruction}

\name{
\begin{tabular}{c}
Chunyu Qiang$^{1,2}$,
Kang Yin$^{2}$, 
Xiaopeng Wang$^{2}$, 
Yuzhe Liang$^{2}$, 
Jiahui Zhao$^{1}$, 
Ruibo Fu$^{3}$, \\ \em 
Tianrui Wang$^{1}$, 
Cheng Gong$^{1}$, 
Chen Zhang$^{2}$, 
Longbiao Wang$^{1}$\thanks{$^\dag$Corresponding Author.},
Jianwu Dang$^{1}$
\end{tabular}
}

\address{
$^1$ Tianjin University, Tianjin, China\\
$^2$  Kuaishou Technology, Beijing, China\\
$^3$  Institute of Automation, Chinese Academy of Sciences, Beijing, China
}

\begin{document}

\ninept
\maketitle

\begin{abstract}
Text-to-speech (TTS) and text-to-music (TTM) models face significant limitations in instruction-based control. TTS systems usually depend on reference audio for timbre, offer only limited text-level attribute control, and rarely support dialogue generation. 
TTM systems are constrained by input conditioning requirements that depend on expert knowledge annotations. The high heterogeneity of these input control conditions makes them difficult to joint modeling with speech synthesis.
Despite sharing common acoustic modeling characteristics, these two tasks have long been developed independently, leaving open the challenge of achieving unified modeling through natural language instructions.
We introduce InstructAudio, a unified framework that enables instruction-based (natural language descriptions) control of acoustic attributes including timbre (gender, age), paralinguistic (emotion, style, accent), and musical (genre, instrument, rhythm, atmosphere). It supports expressive speech, music, and dialogue generation in English and Chinese. 
The model employs joint and single diffusion transformer layers with a standardized instruction-phoneme input format, trained on 50K hours of speech and 20K hours of music data, enabling multi-task learning and cross-modal alignment.
Fig. \ref{fig:radar_comparison} visualizes performance comparisons with mainstream TTS and TTM models, demonstrating that InstructAudio achieves optimal results on most metrics.
To our best knowledge, InstructAudio represents the first instruction-controlled framework unifying speech and music generation. Audio samples are available at: \url{https://qiangchunyu.github.io/InstructAudio/} 
\end{abstract}

%


\begin{figure}[t]
  \centering
  \includegraphics[width=\linewidth]{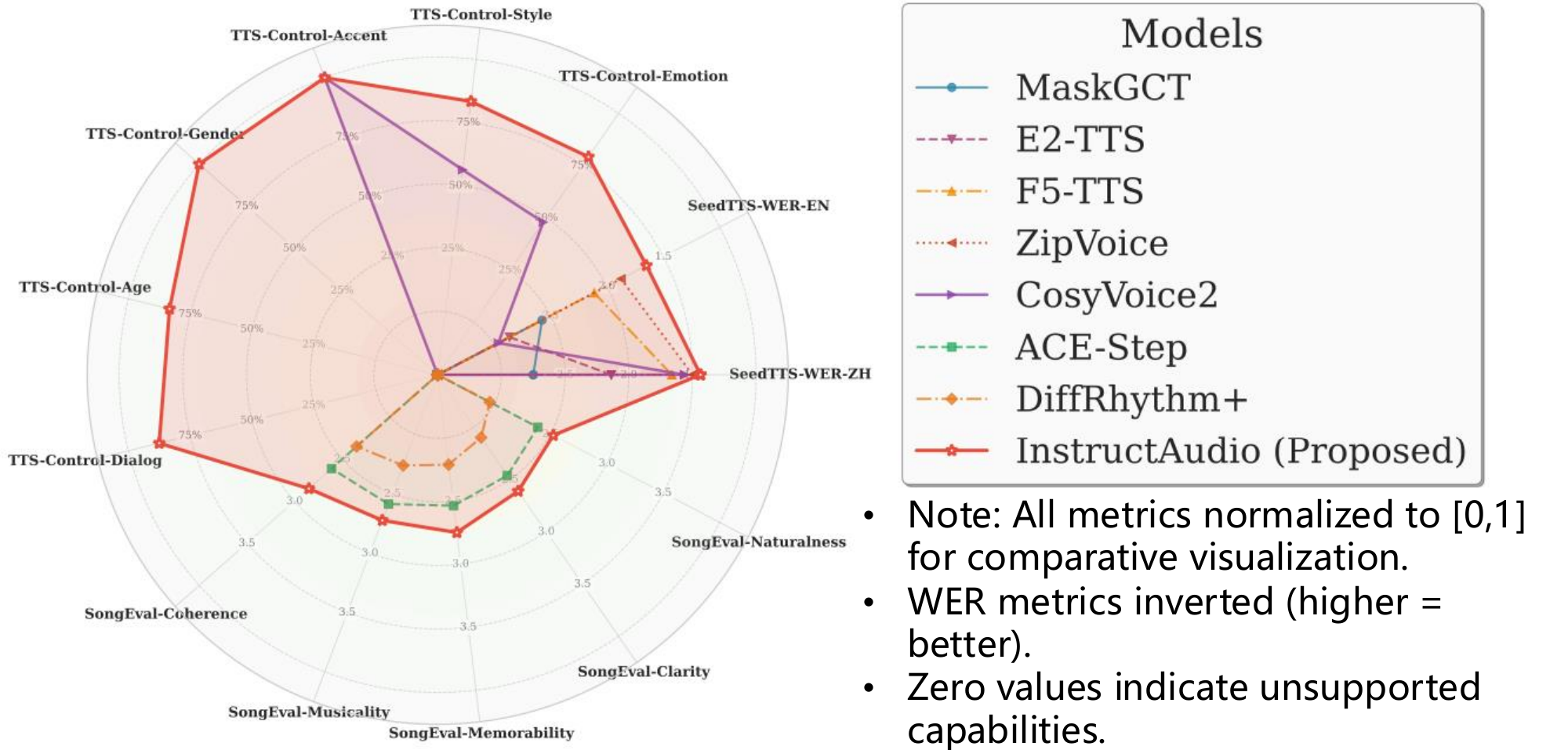}
  \vspace{-0.15in}
  \caption{Comparing model capabilities across TTS and TTM tasks. The chart shows normalized performance on 13 metrics: SeedTTS-WER \cite{anastassiou2024seed}, TTS-Control, and SongEval \cite{yao2025songeval}. InstructAudio (red line) uniquely supports all evaluation dimensions, demonstrating best performance in both TTS and TTM while providing comprehensive controllability across multiple attributes.}
  \vspace{-0.2in}
  \label{fig:radar_comparison}
\end{figure}

\section{Introduction}
\label{sec:intro}

Text-based controllable generation of speech and music is an important research topic in the field of audio generation. Recent developments in TTS \cite{zhang2023speak,kharitonov2023speak,qiang2024minimally,wang2023neural, qiang2024high,du2024cosyvoice2,chen2024f5,shi24f_interspeech, zhu2025zipvoice,eskimez2024e2, 10889461} have achieved impressive results through zero-shot voice cloning and controllable generation. TTM \cite{copet2023simple,gong2025ace, suno2024, udio2024, chen2025diffrhythm+} has advanced with models such as MusicGen \cite{copet2023simple} and ACE-Step \cite{gong2025ace}, alongside commercial systems like Suno\cite{suno2024} and Udio\cite{udio2024}. Despite these advances, instruction-controlled speech and music generation remains a challenging problem in audio processing.

Existing TTS models excel at either zero-shot voice cloning \cite{wang2023neural,du2024cosyvoice,du2024cosyvoice2,chen2024f5} or style control\cite{ji2024controlspeech, guo2023prompttts}, but lack text-based control over multiple acoustic attributes through natural language descriptions. For instance, while CosyVoice \cite{du2024cosyvoice, du2024cosyvoice2} and ControlSpeech \cite{ji2024controlspeech} support text-based emotion and style control, they require additional reference audio for timbre attributes and cannot handle text-controlled dialogue generation. Similarly, current TTM models exhibit limited control capabilities. DiffRhythm+\cite{chen2025diffrhythm+} supports text-based control of genre, instrument, rhythm, and atmosphere but lacks singer timbre control (e.g., gender and age). ACE-Step \cite{gong2025ace}, one of the state-of-the-art (SOTA) open-source music generation model, provides text-based control for all acoustic attributes but focuses exclusively on music without unified speech modeling capabilities.
Speech and music generation are typically treated as separate tasks (TTS \& TTM), overlooking their shared acoustic modeling abilities and control mechanisms. 
This separation stems from the difficulty of aligning inputs across TTS and TTM tasks, as speech control involves acoustic attributes such as timbre and paralinguistics, while music generation requires musical attributes such as genre, instrumentation, and rhythm.

Vevo2 \cite{zhang2025vevo2} introduced the first unified speech and singing generation framework, demonstrating that joint modeling leverages rich speech data to improve singing quality while utilizing singing's expressive characteristics to enhance TTS. However, Vevo2 relies on reference audio for acoustic attribute control rather than text instructions and generates only vocals without instrumental music capabilities.
UniAudio \cite{yang2023uniaudio} builds upon the VALL-E \cite{wang2023neural} framework to create a single model capable of executing multiple tasks; however, it requires inconsistent input formats across different tasks and necessitates task-specific fine-tuning. AudioBox \cite{vyas2023audiobox}, a flow-matching-based unified model supporting multiple tasks, pre-trains on speech, music, and sound effect data but ultimately supports only speech and sound effect generation. AudioLDM 2 \cite{liu2024audioldm} proposes a two-stage model applicable to speech, sound, and music generation, yet requires different model architecture hyperparameters for each task.
Current approaches lack text-based (natural language descriptions) control mechanisms, limiting their ability to achieve unified speech and music generation. To address these limitations, we propose InstructAudio, an instruction-controlled unified framework for speech and music generation.

Our method makes three key contributions:
a) We introduce a joint modeling framework for speech and music generation based on a multimodal diffusion transformer (MM-DiT) architecture.
b) We achieve unified text-based control over TTS and TTM through natural language descriptions (a standardized instruction-phoneme input format), encompassing timbre attributes (gender, age), paralinguistic attributes (emotion, style, accent), and musical attributes (genre, instrument, rhythm, atmosphere), while supporting dialogue speech generation.
c) Experimental results demonstrate best performance in instruction-based TTS (best WER, speaker similarity, emotion similarity, classification control accuracy, and distortion/error metrics) while maintaining competitive TTM capabilities (best SongEval metrics), validating the effectiveness of our unified approach.

\begin{figure*}[t]
 \centering
 \includegraphics[width=0.95\linewidth]{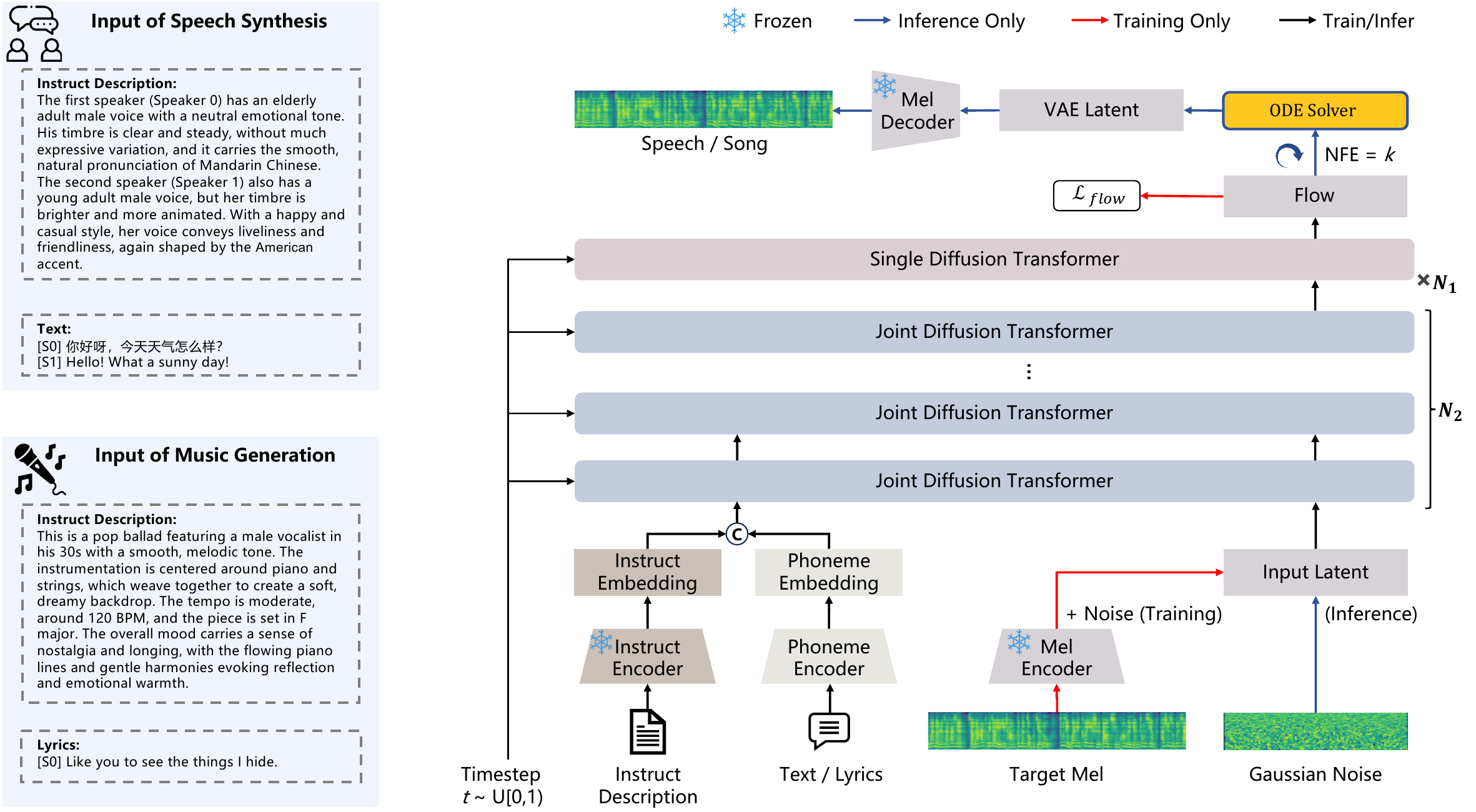}
 \caption{InstructAudio achieves unified generation of both speech and music through an MM-DiT architecture. This framework enables multi-attribute control through natural language instructions. The input format remains consistent across different tasks, comprising a natural language instruction description along with corresponding text or lyrics. Audio is represented using continuous latents extracted from a pre-trained Mel-VAE. During inference, the VAE latent of the target speech or music is obtained through an ODE solver.
 }
 \vspace{-0.15in}
 \label{fig:proposed_model}
\end{figure*}

\section{Method}
\label{sec:method}

\subsection{Overview}
As illustrated in Fig. \ref{fig:proposed_model}, InstructAudio introduces a novel approach for unified speech and music generation, drawing inspiration from MM-Audio \cite{cheng2025mmaudio} through its MM-DiT architecture. The model comprises two core components: joint diffusion transformer layers and single diffusion transformer layers. 
During training, the instruct encoder, mel-encoder, and mel-decoder remain frozen as pretrained modules. The first-layer joint diffusion transformer takes temporally concatenated instruct embedding and phoneme embedding as text modal conditioning input, and noised mel vae latents as audio modal input (see Section \ref{mmdit}).
Unlike existing non-autoregressive architectures\cite{chen2024f5,zhu2025zipvoice,eskimez2024e2}, InstructAudio eliminates the need for text-upsampling alignment with audio representations. For TTS, the model achieves multi-attribute control, including gender, age, emotion, style, and accent, through natural language instructions, and also supports two-speaker dialogue generation. In TTM, it similarly enables multi-attribute control covering singer timbre (e.g., gender and age), music genre, instrumentation, melody, and emotional expression.

\subsection{Unified Instruction-Guided Input}
To enable the unified generation of speech and singing, we designed a standardized instruction-phoneme input format that aligns both tasks, as illustrated in Fig. \ref{fig:proposed_model}. This format consists of two primary components: an instruction description and a phoneme sequence. The instruction description is a natural language prompt specifying the desired acoustic attributes of the output. For speech synthesis, these attributes include speaker characteristics such as gender, age, emotion, style, and accent. In dialogue scenarios, we provide separate descriptions for each of the two speakers and prepend special tokens, [S0] and [S1], to their respective text inputs to differentiate the utterances. Similarly, for music generation, the description specifies attributes like the singer's gender and age, music genre, instrumentation, melody, and emotion. For both tasks, the input text (for speech) or lyrics (for music) is converted into phonemes using a Grapheme-to-Phoneme (G2P) \cite{qiang2022back} model. 
This unified representation allows a single model to seamlessly process input for both speech and music generation. 

\subsection{Latent Audio Codec}
The Latent Audio Codec extends our previous SecoustiCodec framework \cite{qiang2025secousticodec, qiang2025vq, qiang2024learning} and comprises three core components: a mel-encoder, mel-decoder, and discriminator. The VAE architecture enables the model to learn continuous and complete distributions in the latent space, significantly enhancing audio representation capabilities. This high compression ratio facilitates efficient MM-DiT training while improving reconstruction quality for both speech and music. Due to space constraints, we omit the analysis of audio representation effects from this work. For comprehensive results, readers are referred to Section 4.5 of our prior work on Kling-Foley\cite{wang2025kling}.

\subsection{Multimodal Diffusion Transformer Architecture}
\label{mmdit}
We employ conditional flow matching\cite{lipman2022flow} with an MM-DiT architecture based on Stable Diffusion 3\cite{esser2024scaling}, comprising $N_2$ Joint Diffusion Transformer layers. To enhance speech and singing voice generation quality, we incorporate $N_1$ additional Single Diffusion Transformer layers for audio-only processing. The input instruction embeddings $\in \mathbb{R}^{B \times L_1 \times D}$ and phoneme embeddings $\in \mathbb{R}^{B \times L_2 \times D}$ are temporally concatenated to form the text modality input $C_{\text{text}} \in \mathbb{R}^{B \times (L_1+L_2) \times D}$. The linear interpolation path $x_t$ between Gaussian noise and VAE latents serves as the audio modality input.
The two modalities interact through joint attention, where queries, keys, and values from both modalities are concatenated and processed using scaled dot-product attention. The output maintains input dimensionality and is split back into respective modalities. In Single Diffusion Transformer layers, only audio latents are processed, reducing joint attention to self-attention.
During training, flow matching optimizes the objective: $\mathbb{E} \big\| v_\theta(t, C_{\text{text}}, x_t) - u(t, x_t) \big\|^2$
where $v_\theta$ represents the learned conditional velocity field, $u$ denotes the target conditional vector field, and $t$ is the time step. During inference, an ODE solver generates the target VAE latents.

\section{Experiments}

\subsection{Model Details and Datasets}
\label{sec:ModelDetails}
InstructAudio comprises 1.34 billion parameters with a flow matching feedforward dimension of 1024. The architecture includes 14 joint diffusion transformer layers and 6 single diffusion transformer layers, incorporating RoPE positional encoding \cite{su2024roformer}. We employ a Zipformer-based\cite{zhu2025zipvoice} phoneme encoder with a feedforward dimension of 512 and utilize Qwen2.5-7B\cite{qwen2025qwen25technicalreport} as the instruct encoder. The mel encoder processes 44.1kHz input waveforms and generates embeddings at 43 Hz, achieving 1024× downsampling relative to the input sampling rate. Training is conducted on 32 NVIDIA Tesla A800 80GB GPUs with a batch size of 16 per GPU, using the Adam optimizer \cite{Kingma2014AdamAM} with an initial learning rate of 1e-4.

We collect 50K hours of speech and 20K hours of music data from internet sources, applying our internal data processing pipeline to generate instruction descriptions and text/lyrics annotations. For speech data, descriptions encompass gender, age, emotion, style, and accent attributes. Music data descriptions include genre, instrument, gender, age, rhythm, and atmosphere characteristics. Audio clips (2-20s) maintain 1:1 Chinese-English and male-female ratios, with 90\%+ neutral emotions and 0.5\% dialogue data, standardized to 44.1kHz sampling rate.

\subsection{Compared Method and Evaluation Metrics}
\label{sec:ComparedMethod}
To evaluate our model's unified speech and music generation capabilities, we compare against SOTA models in each task. For TTS, we benchmark fundamental capabilities against MaskGCT\cite{wang2024maskgct}, E2-TTS\cite{eskimez2024e2}, F5-TTS\cite{chen2024f5}, ZipVoice\cite{zhu2025zipvoice}, CosyVoice1\cite{du2024cosyvoice}, and CosyVoice2\cite{du2024cosyvoice2} (Table \ref{tab:tts1}), and instruction-based TTS performance against the current SOTA model CosyVoice2 (Table \ref{tab:tts2}). 
Notably, since InstructAudio is a purely instruction-controlled model while the Seed-TTS benchmark comprises natural emotional speech samples, we evaluate WER metrics using natural, calm-style text descriptions with randomized speakers as control conditions. For CosyVoice2, which lacks text-based control for timbre attributes (gender, age), we provide reference audio that matches timbre description during inference and map inputs to its supported short-form control text (e.g., "Please speak very happy").
For music generation, we compare with DiffRhythm+\cite{chen2025diffrhythm+} and ACE-Step\cite{gong2025ace} (Table \ref{tab:music}). Note that DiffRhythm+ does not support music synthesis under 90 seconds; therefore, we generate longer sequences and truncate them to create test samples, which may introduce evaluation bias.

\begin{table}[t]
\centering
\caption{Comparison of TTS models on instruction control and word error rate performance.}
\label{tab:tts1}
\resizebox{\linewidth}{!}{ 
\begin{tabular}{lcccccccc}
\toprule
\multirow{2}{*}{\textbf{Model}} & \multirow{2}{*}{\begin{tabular}[c]{@{}c@{}}\textbf{Data}\textbf{(hrs)}\end{tabular}} & \multirow{2}{*}{\begin{tabular}[c]{@{}c@{}}\textbf{Params}\end{tabular}} & \multicolumn{3}{c}{\textbf{Text-Control}} & \multicolumn{2}{c}{\textbf{WER(\%)$\downarrow$}\cite{anastassiou2024seed}} \\
\cmidrule(lr){4-6} \cmidrule(lr){7-8}
& & & \textbf{G\&A} & \textbf{E\&S\&A} & \textbf{Dial} & \textbf{EN} & \textbf{ZH} \\
\midrule
Ground Truth & -- & -- & -- & -- & -- & 2.14 & 1.25 \\
MaskGCT\cite{wang2024maskgct} & 100K Speech & 1B & \ding{55} & \ding{55} & \ding{55} & 2.26 & 2.40 \\
E2-TTS\cite{eskimez2024e2} & 100K Speech & 333M & \ding{55} & \ding{55} & \ding{55} & 2.49 & 1.91 \\
F5-TTS\cite{chen2024f5} & 100K Speech & 336M & \ding{55} & \ding{55} & \ding{55} & 1.89 & 1.53 \\
ZipVoice\cite{zhu2025zipvoice} & 100K Speech & 123M & \ding{55} & \ding{55} & \ding{55} & 1.70 & 1.40 \\
CosyVoice1\cite{du2024cosyvoice} & 170K Speech & 416M & \ding{55} & \ding{51} & \ding{55} & 4.29 & 3.63 \\
CosyVoice2\cite{du2024cosyvoice2} & 167K Speech & 618M & \ding{55} & \ding{51} & \ding{55} & 2.57 & 1.45 \\
\textbf{InstructAudio} & \begin{tabular}[c]{@{}c@{}}50K Speech \\ + 20K Music\end{tabular} & 1.3B & \ding{51} & \ding{51} & \ding{51} & \textbf{1.52} & \textbf{1.35} \\
\bottomrule
\end{tabular}
}
\begin{tablenotes} 
      \scriptsize
      \item \textbf{Note:} G\&A = Gender\&Age, E\&S\&A = Emotion\&Style\&Accent, Dial = Dialog.
  \end{tablenotes}
\vspace{-0.3in}
\end{table}

\begin{table*}
\centering
\caption{Performance comparison of instruction-based TTS on control accuracy, similarity, distortion/error metrics, and subjective evaluation.}
\label{tab:tts2}
\resizebox{\textwidth}{!}{ 
\begin{tabular}{lccccccccccccccc}
\toprule
\multirow{2}{*}{\textbf{Model}} & \multicolumn{6}{c}{\textbf{Classification Control Accuracy Rate (\%)$\uparrow$}} & \multicolumn{2}{c}{\textbf{Similarity$\uparrow$}} & \multicolumn{4}{c}{\textbf{Distortion/Error $\downarrow$}} & \multicolumn{2}{c}{\textbf{MOS$\uparrow$}} \\
\cmidrule(lr){2-7} \cmidrule(lr){8-9} \cmidrule(lr){10-13} \cmidrule(lr){14-15}
& \textbf{Gender} & \textbf{Age} & \textbf{Emotion} & \textbf{Style} & \textbf{Accent} & \textbf{Dialog} & \textbf{Speaker} & \textbf{Emotion} & \textbf{LSD} & \textbf{MCD} & \textbf{MSEP} & \textbf{MR} & \textbf{QMOS} & \textbf{NMOS} \\
\midrule
Ground Truth & 100.00 & 100.00 & 100.00 & 100.00 & 100.00 & 100.00 & 1.00 & 1.00 & 0.00 & 0.00 & 0.00 & 0.00 & -- & -- \\
CosyVoice2\cite{du2024cosyvoice2} & -- & -- & 58.33 & 65.00 & \textbf{100.00} & -- & 0.68 & 0.53 & 2.57 & 7.11 & 547.87 & 0.46 & \textbf{3.90 ± 0.11} & \textbf{3.65 ± 0.22} \\
\textbf{InstructAudio} & \textbf{100.00} & \textbf{86.67} & \textbf{83.33} & \textbf{86.67} & \textbf{100.00} & \textbf{90.00} & \textbf{0.76} & \textbf{0.71} & \textbf{1.88} & \textbf{5.71} & \textbf{437.58} & \textbf{0.33} & 3.73 ± 0.24 & 3.46 ± 0.32 \\
\bottomrule
\end{tabular}
}
\vspace{-0.1in}
\end{table*}

\begin{table*}[htbp]
\centering
\caption{Performance comparison of TTM on control accuracy, SongEval, and subjective evaluation.}
\label{tab:music}
\resizebox{\textwidth}{!}{ 
\begin{tabular}{lcccccccccccccccc}
\toprule
\multirow{2}{*}{\textbf{Model}} & \multirow{2}{*}{\begin{tabular}[c]{@{}c@{}}\textbf{Data}\textbf{(hrs)}\end{tabular}} & \multirow{2}{*}{\begin{tabular}[c]{@{}c@{}}\textbf{Params}\end{tabular}} & \multicolumn{6}{c}{\textbf{Classification Control Accuracy Rate (\%)$\uparrow$}} & \multicolumn{5}{c}{\textbf{SongEval$\uparrow$\cite{yao2025songeval}}} & \multicolumn{2}{c}{\textbf{MOS$\uparrow$}} \\
\cmidrule(lr){4-9} \cmidrule(lr){10-14} \cmidrule(lr){15-16}
& & & \textbf{Genre} & \textbf{Instrument} & \textbf{Gender} & \textbf{Age} & \textbf{Rhythm} & \textbf{Atmosphere} & \textbf{Coh} & \textbf{Mus} & \textbf{Mem} & \textbf{Cla} & \textbf{Nat} & \textbf{QMOS} & \textbf{MMOS} \\
\midrule
Ground Truth & -- & -- & 100.00 & 100.00 & 100.00 & 100.00 & 100.00 & 100.00 & 3.60 & 3.52 & 3.56 & 3.43 & 3.34 & -- & -- \\
DiffRhythm+\cite{chen2025diffrhythm+} & 120K Music & 1B & 51.33 & 81.67 & 22.22 & 44.44 & 93.33 & 87.22 & 2.68 & 2.61 & 2.57 & 2.48 & 2.37 & 3.04 ± 0.46 & 2.79 ± 0.54 \\
ACE-Step\cite{gong2025ace} & 100K Music & 3B & \textbf{94.44} & \textbf{85.56} & 96.11 & 95.00 & 89.44 & 90.56 & 2.89 & 2.87 & 2.83 & 2.77 & 2.71 & \textbf{3.30 ± 0.28} & 2.88 ± 0.20 \\
\textbf{InstructAudio} & \begin{tabular}[c]{@{}c@{}}50K Speech\\+ 20K Music\end{tabular} & 1.3B & 92.78 & 83.89 & \textbf{98.89} & \textbf{97.22} & \textbf{94.44} & \textbf{95.00} & \textbf{3.08} & \textbf{2.98} & \textbf{3.00} & \textbf{2.89} & \textbf{2.82} & 2.82 ± 0.26 & \textbf{2.91 ± 0.35} \\
\bottomrule
\end{tabular}
}

\begin{tablenotes} 
      \scriptsize
      \item \textbf{Note:} Coh = Coherence, Mus = Musicality, Mem = Memorability, Cla = Clarity, Nat = Naturalness.
  \end{tablenotes}

\vspace{-0.2in}
\end{table*}

We employ comprehensive objective and subjective metrics to ensure thorough evaluation. Objective metrics include Word Error Rate (WER) using Seed-TTS\cite{anastassiou2024seed}, Speaker Similarity\footnote{\url{https://github.com/resemble-ai/Resemblyzer}}, Emotion Similarity\footnote{\url{https://huggingface.co/emotion2vec}}, Log-Spectral Distance (LSD), Mel-Cepstral Distortion (MCD), Mean Squared Error of Pitch (MSEP), Voiced/Unvoiced Mismatch Rate (MR), SongEval\cite{yao2025songeval} music evaluation benchmark, and classification control accuracy through perceptual consistency assessment. Subjective evaluation employs Quality Mean Opinion Score (QMOS), Naturalness Mean Opinion Score (NMOS), and Musicality Mean Opinion Score (MMOS), conducted by professionally trained evaluators. Classification control accuracy is evaluated through human listening tests, where annotators select the perceptually consistent category from predefined options (e.g., choosing among "child," "young/middle-aged," or "elderly" for age attributes).
For WER evaluation, we utilize the complete Seed-TTS benchmark test set. For instruction-based TTS tasks, we construct a manually annotated test set of 500 samples with natural language descriptions covering multiple attributes (gender, age, emotion, style, accent), selecting 100 samples for subjective evaluation. Similarly, for music tasks, we create a 500-sample test set with natural language descriptions of musical attributes (gender, age, genre, instrumentation, melody, emotion), with 100 samples selected for subjective evaluation. The distribution across categories within each attribute is uniform. Notably, InstructAudio's conditioning input consists of descriptive text corresponding to ground truth, with similarity and distortion/error calculations performed against ground truth.

\subsection{Results and Analysis}
\textbf{Evaluation of TTS:} Table \ref{tab:tts1} compares InstructAudio with mainstream TTS models. MaskGCT, E2-TTS, F5-TTS, and ZipVoice do not support text-based control capabilities. CosyVoice1 and CosyVoice2 only support emotion, style, and accent control, requiring additional prompt speech for timbre control. In contrast, InstructAudio supports comprehensive text-based control including gender, age, emotion, style, and accent, while uniquely enabling text-controlled dialogue synthesis (a capability absent in other models). Although InstructAudio has the largest parameter count (1.3B) among TTS models, it additionally supports music generation. As shown in Table \ref{tab:music}, mainstream music generation models like DiffRhythm+ and ACE-Step also exceed 1B parameters. Notably, InstructAudio achieves superior performance with the smallest training dataset. On the Seed-TTS WER metric, InstructAudio achieves the best results when conditioned on neutral emotion and calm style text control.
\textbf{Control Capability:} Table \ref{tab:tts2} evaluates text-based control TTS capabilities by comparing against CosyVoice2, the current SOTA model. For classification control accuracy, InstructAudio supports gender, age, and dialogue control capabilities unavailable in CosyVoice2 (while outperforming CosyVoice2 across all control categories). InstructAudio achieves precise dialogue synthesis control, attaining 90\% accuracy in a capability that other models lack entirely. InstructAudio also demonstrates superior speaker and emotion similarity. This advantage stems from CosyVoice2's reliance on additional prompt speech (random speaker audio with matching gender and age) for timbre control, which causes emotion leakage that compromises emotion control effectiveness. 
InstructAudio outperforms CosyVoice2 across all distortion and error metrics. While CosyVoice2 achieves a higher MOS score, it notably requires reference audio as input conditioning. Text-only control introduces TTS one-to-many mapping ambiguity, reducing average audio quality and naturalness for InstructAudio outputs. 
However, InstructAudio achieves comparable MOS results despite the unfair disadvantage of missing input modality, while comprehensively outperforming CosyVoice2 on all other metrics.
\textbf{Evaluation of TTM:} 
Table \ref{tab:music} evaluates music generation capabilities against current SOTA models ACE-Step and DiffRhythm+. For classification control accuracy, ACE-Step achieves the best performance in genre and instrument categories, while InstructAudio excels in gender, age, rhythm, and atmosphere control. DiffRhythm+ scores poorly on gender and age due to its lack of singer timbre control capabilities. InstructAudio achieves the best SongEval scores across all metrics. ACE-Step obtains the highest QMOS score, indicating superior perceived audio quality, while InstructAudio achieves the best MMOS score. We acknowledge that our music evaluation uses 5-20 second clips matching speech durations, which may disadvantage models like ACE-Step and DiffRhythm+ that are optimized for longer music generation. This comparison demonstrates that InstructAudio maintains competitive music generation capabilities while achieving SOTA text-based control TTS performance.

\subsection{Discussion}
While the text-only control mechanism achieves unified input formatting for TTS and TTM
tasks, it introduces inherent information loss compared to audio modalities, resulting in one-to-many mapping ambiguity. This leads to averaged audio quality and naturalness compared to reference audio-based methods, as evidenced by lower NMOS scores in both TTS and TTM tasks. 
Additionally, to enable joint speech-music modeling, we constrain music generation to 5-20 second clips, limiting long-form music generation capabilities. Our evaluation focuses on short segments to match speech durations, which is detrimental to models like DiffRhythm+ that are optimized for full-length music generation. These limitations above, we expect to address in future work.

\section{Conclusions and future work}

This paper presents InstructAudio, an instruction-controlled unified framework for speech and music generation based on a MM-DiT architecture, demonstrating the effectiveness of joint TTS and TTM modeling. Our main contributions include: (1) introducing the first instruction-controlled unified framework for speech and music generation that eliminates reference audio requirements in attribute control; (2) achieving comprehensive controllability over timbre, paralinguistic, and musical attributes through standardized instruction-phoneme input formatting (natural language descriptions); and (3) demonstrating SOTA performance in instruction-based TTS while maintaining competitive music generation capabilities. Experimental validation confirms our approach's effectiveness, achieving optimal results across multiple metrics including WER, similarity measures, and classification control accuracy. InstructAudio demonstrates the viability of unified audio generation frameworks. Future work will explore multimodal control mechanisms to address quality and naturalness issues, support longer music generation, and incorporate sound effect generation.

\vfill\pagebreak

\scriptsize
\bibliographystyle{IEEEbib}
\bibliography{refs}

@article{qiang2025secousticodec,
  title={SecoustiCodec: Cross-Modal Aligned Streaming Single-Codecbook Speech Codec},
  author={Qiang, Chunyu and Wang, Haoyu and Gong, Cheng and Wang, Tianrui and Fu, Ruibo and Wang, Tao and Chen, Ruilong and Yi, Jiangyan and Wen, Zhengqi and Zhang, Chen and others},
  journal={arXiv preprint arXiv:2508.02849},
  year={2025}
}

@article{yao2025songeval,
  title={SongEval: A Benchmark Dataset for Song Aesthetics Evaluation},
  author={Yao, Jixun and Ma, Guobin and Xue, Huixin and Chen, Huakang and Hao, Chunbo and Jiang, Yuepeng and Liu, Haohe and Yuan, Ruibin and Xu, Jin and Xue, Wei and others},
  journal={arXiv preprint arXiv:2505.10793},
  year={2025}
}

@article{qwen2025qwen25technicalreport,
      title={Qwen2.5:Technical Report}, 
      author={An Yang and Baosong Yang and Beichen Zhang and Binyuan Hui and Bo Zheng, et al.},
      year={2025},
      eprint={2412.15115},
      archivePrefix={arXiv},
      primaryClass={cs.CL},
      url={https://arxiv.org/abs/2412.15115}, 
}

@article{su2024roformer,
  title={Roformer: Enhanced transformer with rotary position embedding},
  author={Su, Jianlin and Ahmed, Murtadha and Lu, Yu and Pan, Shengfeng and Bo, Wen and Liu, Yunfeng},
  journal={Neurocomputing},
  volume={568},
  pages={127063},
  year={2024},
  publisher={Elsevier}
}

@inproceedings{esser2024scaling,
  title={Scaling rectified flow transformers for high-resolution image synthesis},
  author={Esser, Patrick and Kulal, Sumith and Blattmann, Andreas and Entezari, Rahim and M{\"u}ller, Jonas and Saini, Harry and Levi, Yam and Lorenz, Dominik and Sauer, Axel and Boesel, Frederic and others},
  booktitle={Forty-first international conference on machine learning},
  year={2024}
}

@article{vyas2023audiobox,
  title={Audiobox: Unified audio generation with natural language prompts},
  author={Vyas, Apoorv and Shi, Bowen and Le, Matthew and Tjandra, Andros and Wu, Yi-Chiao and Guo, Baishan and Zhang, Jiemin and Zhang, Xinyue and Adkins, Robert and Ngan, William and others},
  journal={arXiv preprint arXiv:2312.15821},
  year={2023}
}

@article{yang2023uniaudio,
  title={Uniaudio: An audio foundation model toward universal audio generation},
  author={Yang, Dongchao and Tian, Jinchuan and Tan, Xu and Huang, Rongjie and Liu, Songxiang and Chang, Xuankai and Shi, Jiatong and Zhao, Sheng and Bian, Jiang and Wu, Xixin and others},
  journal={arXiv preprint arXiv:2310.00704},
  year={2023}
}

@article{liu2024audioldm,
  title={Audioldm 2: Learning holistic audio generation with self-supervised pretraining},
  author={Liu, Haohe and Yuan, Yi and Liu, Xubo and Mei, Xinhao and Kong, Qiuqiang and Tian, Qiao and Wang, Yuping and Wang, Wenwu and Wang, Yuxuan and Plumbley, Mark D},
  journal={IEEE/ACM Transactions on Audio, Speech, and Language Processing},
  volume={32},
  pages={2871--2883},
  year={2024},
  publisher={IEEE}
}

@article{lipman2022flow,
  title={Flow matching for generative modeling},
  author={Lipman, Yaron and Chen, Ricky TQ and Ben-Hamu, Heli and Nickel, Maximilian and Le, Matt},
  journal={arXiv preprint arXiv:2210.02747},
  year={2022}
}

@article{wang2025kling,
  title={Kling-Foley: Multimodal Diffusion Transformer for High-Quality Video-to-Audio Generation},
  author={Wang, Jun and Zeng, Xijuan and Qiang, Chunyu and Chen, Ruilong and Wang, Shiyao and Wang, Le and Zhou, Wangjing and Cai, Pengfei and Zhao, Jiahui and Li, Nan and others},
  journal={arXiv preprint arXiv:2506.19774},
  year={2025}
}

@inproceedings{qiang2022back,
  title={Back-Translation-Style Data Augmentation for Mandarin Chinese Polyphone Disambiguation},
  author={Qiang, Chunyu and Yang, Peng and Che, Hao and Xiao, Jinba and Wang, Xiaorui and Wang, Zhongyuan},
  booktitle={2022 Asia-Pacific Signal and Information Processing Association Annual Summit and Conference (APSIPA ASC)},
  pages={1915--1919},
  year={2022},
  organization={IEEE}
}

@article{wang2023neural,
  title={Neural Codec Language Models are Zero-Shot Text to Speech Synthesizers},
  author={Wang, Chengyi and Chen, Sanyuan and Wu, Yu and Zhang, Ziqiang and Zhou, Long and Liu, Shujie and Chen, Zhuo and Liu, Yanqing and Wang, Huaming and Li, Jinyu and others},
  journal={arXiv preprint arXiv:2301.02111},
  year={2023}
}

@article{kharitonov2023speak,
  title={Speak, Read and Prompt: High-Fidelity Text-to-Speech with Minimal Supervision},
  author={Kharitonov, Eugene and Vincent, Damien and Borsos, Zal{\'a}n and Marinier, Rapha{\"e}l and Girgin, Sertan and Pietquin, Olivier and Sharifi, Matt and Tagliasacchi, Marco and Zeghidour, Neil},
  journal={arXiv preprint arXiv:2302.03540},
  year={2023}
}

@article{zhang2025vevo2,
  title={Vevo2: Bridging Controllable Speech and Singing Voice Generation via Unified Prosody Learning},
  author={Zhang, Xueyao and Zhang, Junan and Wang, Yuancheng and Wang, Chaoren and Chen, Yuanzhe and Jia, Dongya and Chen, Zhuo and Wu, Zhizheng},
  journal={arXiv preprint arXiv:2508.16332},
  year={2025}
}

@article{zhang2023speak,
  title={Speak foreign languages with your own voice: Cross-lingual neural codec language modeling},
  author={Zhang, Ziqiang and Zhou, Long and Wang, Chengyi and Chen, Sanyuan and Wu, Yu and Liu, Shujie and Chen, Zhuo and Liu, Yanqing and Wang, Huaming and Li, Jinyu and others},
  journal={arXiv preprint arXiv:2303.03926},
  year={2023}
}

@article{ji2024controlspeech,
  title={Controlspeech: Towards simultaneous zero-shot speaker cloning and zero-shot language style control with decoupled codec},
  author={Ji, Shengpeng and Zuo, Jialong and Wang, Wen and Fang, Minghui and Zheng, Siqi and Chen, Qian and Jiang, Ziyue and Huang, Hai and Wang, Zehan and Cheng, Xize and others},
  journal={arXiv preprint arXiv:2406.01205},
  year={2024}
}

@inproceedings{guo2023prompttts,
  title={Prompttts: Controllable text-to-speech with text descriptions},
  author={Guo, Zhifang and Leng, Yichong and Wu, Yihan and Zhao, Sheng and Tan, Xu},
  booktitle={ICASSP 2023-2023 IEEE International Conference on Acoustics, Speech and Signal Processing (ICASSP)},
  pages={1--5},
  year={2023},
  organization={IEEE}
}

@article{Kingma2014AdamAM,
  title={Adam: A Method for Stochastic Optimization},
  author={Diederik P. Kingma and Jimmy Ba},
  journal={CoRR},
  year={2014},
  volume={abs/1412.6980}
}

@article{wang2024maskgct,
  title={Maskgct: Zero-shot text-to-speech with masked generative codec transformer},
  author={Wang, Yuancheng and Zhan, Haoyue and Liu, Liwei and Zeng, Ruihong and Guo, Haotian and Zheng, Jiachen and Zhang, Qiang and Zhang, Xueyao and Zhang, Shunsi and Wu, Zhizheng},
  journal={arXiv preprint arXiv:2409.00750},
  year={2024}
}

@article{chen2024f5,
  title={F5-tts: A fairytaler that fakes fluent and faithful speech with flow matching},
  author={Chen, Yushen and Niu, Zhikang and Ma, Ziyang and Deng, Keqi and Wang, Chunhui and Zhao, Jian and Yu, Kai and Chen, Xie},
  journal={arXiv preprint arXiv:2410.06885},
  year={2024}
}

@article{zhu2025zipvoice,
  title={ZipVoice: Fast and High-Quality Zero-Shot Text-to-Speech with Flow Matching},
  author={Zhu, Han and Kang, Wei and Yao, Zengwei and Guo, Liyong and Kuang, Fangjun and Li, Zhaoqing and Zhuang, Weiji and Lin, Long and Povey, Daniel},
  journal={arXiv preprint arXiv:2506.13053},
  year={2025}
}

@inproceedings{eskimez2024e2,
  title={E2 tts: Embarrassingly easy fully non-autoregressive zero-shot tts},
  author={Eskimez, Sefik Emre and Wang, Xiaofei and Thakker, Manthan and Li, Canrun and Tsai, Chung-Hsien and Xiao, Zhen and Yang, Hemin and Zhu, Zirun and Tang, Min and Tan, Xu and others},
  booktitle={2024 IEEE Spoken Language Technology Workshop (SLT)},
  pages={682--689},
  year={2024},
  organization={IEEE}
}

@inproceedings{qiang2024minimally,
  title={Minimally-supervised speech synthesis with conditional diffusion model and language model: A comparative study of semantic coding},
  author={Qiang, Chunyu and Li, Hao and Ni, Hao and Qu, He and Fu, Ruibo and Wang, Tao and Wang, Longbiao and Dang, Jianwu},
  booktitle={ICASSP 2024-2024 IEEE International Conference on Acoustics, Speech and Signal Processing (ICASSP)},
  pages={10186--10190},
  year={2024},
  organization={IEEE}
}

@inproceedings{qiang2024learning,
  title={Learning speech representation from contrastive token-acoustic pretraining},
  author={Qiang, Chunyu and Li, Hao and Tian, Yixin and Fu, Ruibo and Wang, Tao and Wang, Longbiao and Dang, Jianwu},
  booktitle={ICASSP 2024-2024 IEEE International Conference on Acoustics, Speech and Signal Processing (ICASSP)},
  pages={10196--10200},
  year={2024},
  organization={IEEE}
}

@inproceedings{qiang2024high,
  title={High-Fidelity Speech Synthesis with Minimal Supervision: All Using Diffusion Models},
  author={Qiang, Chunyu and Li, Hao and Tian, Yixin and Zhao, Yi and Zhang, Ying and Wang, Longbiao and Dang, Jianwu},
  booktitle={ICASSP 2024-2024 IEEE International Conference on Acoustics, Speech and Signal Processing (ICASSP)},
  pages={10781--10785},
  year={2024},
  organization={IEEE}
}

@inproceedings{cheng2025mmaudio,
  title={MMAudio: Taming Multimodal Joint Training for High-Quality Video-to-Audio Synthesis},
  author={Cheng, Ho Kei and Ishii, Masato and Hayakawa, Akio and Shibuya, Takashi and Schwing, Alexander and Mitsufuji, Yuki},
  booktitle={Proceedings of the Computer Vision and Pattern Recognition Conference},
  pages={28901--28911},
  year={2025}
}

@article{qiang2025vq,
  title={VQ-CTAP: Cross-Modal Fine-Grained Sequence Representation Learning for Speech Processing},
  author={Qiang, Chunyu and Geng, Wang and Zhao, Yi and Fu, Ruibo and Wang, Tao and Gong, Cheng and Wang, Tianrui and Liu, Qiuyu and Yi, Jiangyan and Wen, Zhengqi and others},
  journal={IEEE Transactions on Audio, Speech and Language Processing},
  year={2025},
  publisher={IEEE}
}

@article{anastassiou2024seed,
  title={Seed-tts: A family of high-quality versatile speech generation models},
  author={Anastassiou, Philip and Chen, Jiawei and Chen, Jitong and Chen, Yuanzhe and Chen, Zhuo and Chen, Ziyi and Cong, Jian and Deng, Lelai and Ding, Chuang and Gao, Lu and others},
  journal={arXiv preprint arXiv:2406.02430},
  year={2024}
}

@article{copet2023simple,
  title={Simple and controllable music generation},
  author={Copet, Jade and Kreuk, Felix and Gat, Itai and Remez, Tal and Kant, David and Synnaeve, Gabriel and Adi, Yossi and D{\'e}fossez, Alexandre},
  journal={Advances in Neural Information Processing Systems},
  volume={36},
  pages={47704--47720},
  year={2023}
}

@misc{suno2024,
  author = {Suno AI},
  title = {Suno: AI Music Generation Platform},
  year = {2024},
  howpublished = {\url{https://suno.com}}
}

@misc{udio2024,
  author = {Udio AI},
  title = {Udio: AI Music Creation Platform},
  year = {2024},
  howpublished = {\url{https://udio.com}}
}

@article{du2024cosyvoice,
  title={Cosyvoice: A scalable multilingual zero-shot text-to-speech synthesizer based on supervised semantic tokens},
  author={Du, Zhihao and Chen, Qian and Zhang, Shiliang and Hu, Kai and Lu, Heng and Yang, Yexin and Hu, Hangrui and Zheng, Siqi and Gu, Yue and Ma, Ziyang and others},
  journal={arXiv preprint arXiv:2407.05407},
  year={2024}
}

@article{du2024cosyvoice2,
  title={Cosyvoice 2: Scalable streaming speech synthesis with large language models},
  author={Du, Zhihao and Wang, Yuxuan and Chen, Qian and Shi, Xian and Lv, Xiang and Zhao, Tianyu and Gao, Zhifu and Yang, Yexin and Gao, Changfeng and Wang, Hui and others},
  journal={arXiv preprint arXiv:2412.10117},
  year={2024}
}

@article{chen2025diffrhythm+,
  title={DiffRhythm+: Controllable and Flexible Full-Length Song Generation with Preference Optimization},
  author={Chen, Huakang and Jiang, Yuepeng and Ma, Guobin and Hao, Chunbo and Wang, Shuai and Yao, Jixun and Ning, Ziqian and Meng, Meng and Luan, Jian and Xie, Lei},
  journal={arXiv preprint arXiv:2507.12890},
  year={2025}
}

@article{gong2025ace,
  title={Ace-step: A step towards music generation foundation model},
  author={Gong, Junmin and Zhao, Sean and Wang, Sen and Xu, Shengyuan and Guo, Joe},
  journal={arXiv preprint arXiv:2506.00045},
  year={2025}
}

@INPROCEEDINGS{10889461,
  author={Qi, Xin and Fu, Ruibo and et al},
  booktitle={ICASSP 2025 - 2025 IEEE International Conference on Acoustics, Speech and Signal Processing (ICASSP)}, 
  title={DPI-TTS: Directional Patch Interaction for Fast-Converging and Style Temporal Modeling in Text-to-Speech}, 
  year={2025},
  pages={1-5},
  doi={10.1109/ICASSP49660.2025.10889461}}

@inproceedings{shi24f_interspeech,
  title     = {{PPPR: Portable Plug-in Prompt Refiner for Text to Audio Generation}},
  author    = {Shuchen Shi and Ruibo Fu and et al},
  year      = {2024},
  booktitle = {{Interspeech 2024}},
  pages     = {4898--4902},
  doi       = {10.21437/Interspeech.2024-1771},
  issn      = {2958-1796},
}


\end{document}